%% file: main.tex
\documentclass[webpdf,modern,large,namedate]{oup-authoring-template}%

\usepackage[utf8]{inputenc} 
\usepackage[T1]{fontenc}    
\usepackage{url}            
\usepackage{booktabs}       
\usepackage{amsfonts}       
\usepackage{nicefrac}       
\usepackage{microtype}      
\usepackage{xcolor}         
\usepackage{graphicx}
\usepackage{caption}
\usepackage{subcaption}
\usepackage{multirow}
\usepackage{advdate}
\usepackage{amssymb}
\usepackage{amsmath}
\usepackage{mathrsfs} 
\usepackage{systeme}
\usepackage{mathtools}
\usepackage{pdfpages}
\usepackage{paralist}
\usepackage{algorithm}
\usepackage{xspace}
\usepackage{array}
\usepackage{comment}
\usepackage[noend]{algpseudocode}
\usepackage{tikz}
\usetikzlibrary{matrix, positioning, arrows.meta, calc, shapes, decorations, backgrounds, arrows, decorations.pathreplacing, automata}
\usepackage[round]{natbib}
\usepackage{hyperref}%
\hypersetup{colorlinks=true, citecolor=[rgb]{0,0,1}, urlcolor=black}
\usepackage{pgfplots}

\graphicspath{{Fig/}}

\theoremstyle{thmstyleone}%
\theoremstyle{thmstyletwo}%
\theoremstyle{thmstylethree}%

\renewcommand{\cite}[1]{\citep{{#1}}} 
\newcommand{\namecite}[1]{\citet{{#1}}} 

\algdef{SE}[SUBALG]{Indent}{EndIndent}{}{\algorithmicend\ }%
\algtext*{Indent}
\algtext*{EndIndent}

\input{defs}

\begin{document}

\journaltitle{ISMB}
\DOI{DOI HERE}
\copyrightyear{2024}
\pubyear{2019}
\access{Advance Access Publication Date: Day Month Year}
\appnotes{Paper}

\firstpage{1}


\title[mRNA Design via Expected Partition Function 
and Continuous Optimization]
{Messenger RNA Design via Expected Partition Function 
and  Continuous Optimization}

\author[1]{Ning Dai}
\author[1]{Wei Yu Tang}
\author[1]{Tianshuo Zhou}
\author[3,4,5]{David H.~Mathews}
\author[1,2,$\ast$]{Liang Huang}

\authormark{Dai et al.}

\address[1]{\orgdiv{School of Electrical Engineering \& Computer Science}, \orgname{ Oregon State University}, \orgaddress{\street{Corvallis}, \postcode{97330}, \state{OR}, \country{USA}}}
\address[2]{\orgdiv{Dept.~of Biochemistry \& Biophysics}, \orgname{Oregon State University},  \orgaddress{\street{Corvallis}, \postcode{97330}, \state{OR}, \country{USA}}}
\address[3]{\orgdiv{Dept. of Biochemistry \& Biophysics}}
\address[4]{\orgdiv{Center for RNA Biology}}
\address[5]{\orgdiv{Dept. of Biostatistics \& Computational Biology}, \orgname{ University of Rochester Medical Center}, \orgaddress{\street{Rochester}, \postcode{14642}, \state{NY}, \country{14642}}}

\corresp[$\ast$]{Corresponding author. \href{email:liang.huang.sh@gmail.com}{liang.huang.sh@gmail.com}}




\abstract{The tasks of designing 
RNAs are discrete optimization problems, and
several versions of these problems are NP-hard.
As an alternative to commonly used local search methods,
we formulate these problems 
as continuous optimization and develop
a general framework for this optimization based on
a generalization of classical partition function which we call
``expected partition function''.
The basic idea is to start with 
a distribution over all possible candidate sequences,
and extend the objective function from a sequence to a distribution. We then use gradient descent-based optimization methods
to improve the extended objective function, and the distribution 
will gradually shrink towards a one-hot sequence (i.e., a single sequence).
As a case study, we consider the important problem of mRNA design 
with wide applications in vaccines and therapeutics.
While the recent work of LinearDesign can efficiently 
optimize mRNAs for minimum free energy (MFE),
optimizing for ensemble free energy is much harder
and likely intractable.
Our approach can consistently 
improve over the LinearDesign solution
in terms of ensemble free energy,
with bigger improvements on longer sequences.}


\maketitle

\section{Introduction}
\input{introduction}

\section{Optimization Framework and Expected Partition Function}
\label{sec:framework}
\input{methods}

\section{mRNA Design for Minimum Ensemble Free Energy}
\label{sec:mRNA}


\input{mRNA}

\section{Constrained Optimization via Projected Gradient Descent}
\label{sec:optimize}
\input{optimization}

\section{Experiments and Results}
\label{sec:results}

\input{results-mRNA}

\section{Related Work}
\label{sec:related}
\input{related}

\section{Conclusions and Future Work}

\input{conclusions.tex}


\input{fig_LD_partition.tex}

\input{fig_beam_pruning.tex}

\bibliographystyle{natbib}
\bibliography{references}

\newpage

\section*{Appendix: Probabilistic mRNA DFAs}
\input{DFA_formulation.tex}

\end{document}

%% file: defs.tex
\renewcommand{\vec}[1]{\ensuremath{\mathbf{{#1}}}\xspace}

\newcommand{\EFE}{\ensuremath{\freeenergy_{\mathrm{ens}}}\xspace}
\newcommand{\genEFE}{\ensuremath{\widetilde{\freeenergy}_{\mathrm{ens}}}\xspace}
\newcommand{\DeltaEFE}{\ensuremath{\Delta\freeenergy_{\mathrm{ens}}}\xspace}
\newcommand{\aas}{\ensuremath{\text{\it aa}}\xspace}
\newcommand{\nts}{\ensuremath{\text{\it nt}}\xspace}

\newcommand{\notes}[1]{}

\newcommand{\ith}[1]{\ensuremath{i^{{th}}}}

\newcount\permx
\newcount\permy
\def\permdot#1#2{
\permx=#1 \advance\permx by-1
\permy=#2 \advance\permy by-1
\psframe[fillcolor=black, fillstyle=solid]
(\permx,\permy)(#1, #2)
}

\newcommand{\argmin}{\operatornamewithlimits{\mathrm{argmin}}}

\newcommand{\cands}{\ensuremath{\textit{cands}}\xspace}
\newcommand{\nodes}{\ensuremath{\textit{nodes}}\xspace}

\newcommand{\pluseq}{\mathrel{+}=}

\newcommand{\vecx}{\ensuremath{\vec{x}}\xspace}
\newcommand{\vecy}{\ensuremath{\vec{y}}\xspace}

\newcommand{\vecp}{\ensuremath{\vec{p}}\xspace}

\newcommand{\vecX}{\ensuremath{\mathbf{X}}\xspace}
\newcommand{\vectheta}{\ensuremath{\boldsymbol{\theta}}\xspace}

\newcommand{\subscript}[2]{_{{#1}, {#2}}}
\newcommand{\Qf}[2]{\ensuremath{Q\subscript{{#1}}{{#2}}}\xspace}
\newcommand{\Qsf}[3]{\ensuremath{Q^{#1}\subscript{{#2}}{{#3}}}\xspace}
\newcommand{\expQsf}[3]{\ensuremath{\expQ^{#1}\subscript{{#2}}{{#3}}}\xspace}

\newcommand{\smallnt}[1]{\ensuremath{_{\mbox{\tiny PP}}}\xspace}

\iffalse

\else

\fi

\newcommand{\GEN}{\ensuremath{\mathcal{Y}}\xspace}

\newcommand{\smallurl}[1]{{\scriptsize \url{#1}}}

\newcommand{\nucA}{\ensuremath{\text{\sc a}}}
\newcommand{\nucU}{\ensuremath{\text{\sc u}}}
\newcommand{\nucC}{\ensuremath{\text{\sc c}}}
\newcommand{\nucG}{\ensuremath{\text{\sc g}}}

\newcommand{\allpairtypes}{\ensuremath{\{\nucA\nucU, \nucU\nucA, \nucC\nucG, \nucG\nucC, \nucG\nucU, \nucU\nucG\}}\xspace}

\newcommand{\freeenergy}{\ensuremath{\Delta G^\circ}\xspace}
\newcommand{\ensemble}{\ensuremath{\text{ensemble}}\xspace}
\newcommand{\ensenergy}{\EFE}
\newcommand{\genEnsenergy}{\ensuremath{\widetilde{\freeenergy}_\ensemble}\xspace}

\newcommand{\outedges}{\ensuremath{\mathit{out\scalebox{0.5}{\_}edges}}\xspace}
\newcommand{\inedges}{\ensuremath{\mathit{in\scalebox{0.5}{\_}edges}}\xspace}

\definecolor{intnull}{RGB}{213,229,255}
\definecolor{inteins}{RGB}{128,179,255}
\definecolor{intvier}{RGB}{42,127,255}
\definecolor{intdrei}{RGB}{0,85,212}
\definecolor{intvier}{RGB}{0,51,128}
\definecolor{intfunf}{RGB}{0,34,85}

\newcommand{\laedgew}[3]{\ensuremath{\text{\footnotesize \ensuremath{#1}}\stackrel{#2}{\longrightarrow}\text{\footnotesize \ensuremath{#3}}}\xspace} 
\newcommand{\longlaedgew}[4][4em]{%
    \ensuremath{\text{\footnotesize \ensuremath{#2}}\xrightarrow{\makebox[#1]{\tiny \ensuremath{#3}}}\text{\footnotesize \ensuremath{#4}}}\xspace
}

\newsavebox\CBox

\DeclareMathOperator{\E}{\mathbb{E}}

\newcommand{\expQ}{\ensuremath{\widetilde{Q}}\xspace}
\newcommand{\expQy}{\ensuremath{\expQ_{\vecy}}\xspace}
\newcommand{\nucset}{\ensuremath{\mathsf{N}}\xspace}

\newcommand{\distri}{\ensuremath{\mathbb{D}}\xspace}

%% file: introduction.tex

Ribonucleic acid (RNA) is of utmost importance in life because it plays 
both the {\em informational} role where
messenger RNAs (mRNAs) pass genetic information from DNAs to proteins
as well as the {\em functional} roles where 
non-coding RNAs facilitate protein translation, catalyze reactions, and
regulate gene expression. 
By contrast, DNA has an informational role and protein only functional ones.
In addition, the public became more aware of  RNA because COVID-19 was caused by the RNA virus SARS-CoV-2 \cite{wu+:2020:sarscov2}, which was then partially contained by mRNA vaccines \cite{baden+:2021,polack+:2020}.

The field of RNA design has applications in diagnostics and therapeutics
\cite{zhang+:2023,zhou+:2023samfeo}. 
There are two broad categories of RNA design problems, which correspond
to the two roles of RNAs above. 
The first type (``information passing'') is to design optimal mRNAs given a protein, 
which aims to find the optimal mRNA 
in terms of factors such as stability and codon optimality 
among all the mRNA sequences that translate to that protein \cite{mauger+:2019,cohen+skiena:2003,terai+:2016}.
For example, \namecite{zhang+:2023} designed an efficient algorithm for mRNA design that jointly optimizes minimum free energy (MFE) and codon optimality,
by reducing the mRNA design problem to lattice parsing, where the design space is a regular language (encoded by a finite-state automata) and the objective function (energy) is encoded by a stochastic context-free grammar. 
However, the mRNA design problem optimizing for partition function \cite{mccaskill:1990} (i.e., ensemble free energy) is wide open, and is likely NP-hard due to the minimization over a summation (instead of minimization over a minimization in the MFE case).
The second type (structural/functional) 
is to design an non-coding RNA for a secondary structure,
which aims to find an RNA sequence that naturally folds,
or has the highest probability of folding, into that structure.
Some instance of this problem has been proved to be NP-hard \cite{bonnet2020designing}.

While these RNA design problems are by default discrete optimization problems
that were tackled by local search methods such as random walk \cite{zhou+:2023samfeo},
we offer an alternative view and instead formulate them as continuous optimization instances.
The basic idea is to start with 
a distribution over all possible candidate sequences,
and extend the objective function from a sequence to 
be the expectation on a distribution. 
This extension employs
a generalization of the classical single-sequence partition function \cite{mccaskill:1990}
which we call
``expected partition function'' (Section~\ref{sec:framework}).
\namecite{matthies+:2023} proposed this concept
but from a different perspective and for a different task
(see Sec.~\ref{sec:related} 
for details).
We then use gradient descent-based optimization methods (Section~\ref{sec:optimize})
to improve the extended objective function, and the distribution 
will gradually shrink towards a one-hot sequence (i.e., a single sequence).
As a case study, we consider the important problem 
of the mRNA design problem which 
has broad applications in both vaccines and therapeutics (Section~\ref{sec:mRNA}).
While previous work like LinearDesign
can efficiently optimize mRNAs for minimum free energy (MFE),
optimizing for ensemble free energy is a much harder problem
and is likely intractable. 
On 20 protein sequences, our algorithm can consistently improve over LinearDesign's best-MFE results in terms of ensemble free energy,
with bigger improvements on longer sequences.

%
%
%
%

%% file: methods.tex

In our computational representation, an RNA sequence is denoted as $\vecx = (x_1, x_2, .., x_{n})$, with each $x_i \in \nucset$ representing the $i$-th nucleotide of the sequence, drawn from the set of all possible nucleotides, $\nucset= \{\nucA, \nucC, \nucG, \nucU\}$. To encapsulate a distribution over sequences, we introduce individual random variables $X_i$, each corresponding to a specific position in the RNA sequence.  The distribution of \(X_i\) over the nucleotide set is given by a Probability Mass Function (PMF), \(\distri_i\), such that:
\[
\distri_i : \nucset \mapsto [0, 1], \text{ with } \sum_{x_i\in \nucset} \distri_i(x_i) = 1,
\]
which implies that the sum of the probabilities of all possible nucleotides at position \(i\) equals 1.

Here, the observed nucleotide at a given position, $x_i$, is considered as a sample drawn from the associated random variable $X_i$, i.e., $x_i \sim \distri_i(\cdot)$.  Within this framework, an entire RNA sequence, $\vecx$, is conceived as a sample drawn from the joint distribution of all positions, $\vecX$, i.e., $\vecx \sim \distri(\cdot)$. The joint PMF defining this distribution is defined as the product of these individual distributions, represented by \(\distri: \nucset^n \mapsto [0,1]\). It is formulated as:
\[
\distri(\vecx) = \prod_{i=1\ldots n} \distri_i(x_i),
\]
where \(\sum_{\vecx \in \nucset^n} \distri(\vecx) = 1\), ensuring that the total probability over all positions.

\begin{figure*}
\algrenewcommand\algorithmicindent{1em}%
\label{alg:partition}

\centering
\begin{minipage}{.49\textwidth}
\begin{algorithmic}
\Function{Partition}{$\vecx$}
    \State $n \gets$ length of $\vecx$
    \State $Q \gets$ hash() \Comment{hash table: from span $[i,j]$ to $\Qf{i}{j}$}
    \State $\Qf{j}{j-1} \gets 1$ for all $j$ in $1...n$ \Comment{base cases}
    \For{$j = 1$ to $n$}
        \ForAll {$i$ such that $[i,\,j-1]$ in $Q$}
            \State $\Qf{i}{j} \pluseq \Qf{i}{j-1} \cdot e^{-\frac{\delta(x_j)}{RT}} $
            \If{$x_{i-1}x_j$ in \{\nucA\nucU, \nucU\nucA, \nucC\nucG, \nucG\nucC, \nucG\nucU, \nucU\nucG\}}
                \ForAll{$k$ such that $[k,\,i-2]$ in $Q$} 
                    \State $\Qf{k}{j} \pluseq \Qf{k}{i-2} \cdot \Qf{i}{j-1} \cdot e^{-\frac{\xi(x_{i-1},x_j)}{RT}}$
                \EndFor
            \EndIf
        \EndFor
    \EndFor
    \State \Return $Q$ \Comment{Partition function $Q(\vecx)=\Qf{1}{n}$}
\EndFunction
\end{algorithmic}
\end{minipage} 
\begin{minipage}{.49\textwidth}
\label{alg:expected_partition}
\begin{algorithmic}
\Function{ExPartition}{{\color{red}$\distri =   \prod_{i=1}^{n} \distri_i$}} \Comment{{\color{red}$\distri$ is a distribution}}
    \State $\expQ \gets$ hash() \Comment{hash table: from span $[i,j]$ to $\expQ_{i,j}$}
    \State $\expQ_{j,j-1} \gets 1$ for all $j$ in $1...n$ \Comment{base cases}
    \For{$j = 1$ to $n$}
        \ForAll {$i$ such that $[i,\,j-1]$ in $\expQ$}
           \textcolor{red}{
            \ForAll{$c$ in \{\nucA, \nucC, \nucG, \nucU\}} \Comment{all nucs at position $j$}
                \State  $\textcolor{black}{ \expQ_{i,j} \pluseq } \distri_j(c) {\color{black}\, \cdot\, \expQ_{i,j-1} \cdot e^{-\frac{\delta({\color{red}c})}{RT}}} $
            \EndFor
            \ForAll{$a, b$ in \allpairtypes} \Comment{all pairs}           
           	\textcolor{black}{
                \ForAll{$k$ such that $[k,\,i-2]$ in $\expQ$} 
                    \State $\expQ_{k,j} \pluseq \textcolor{red}{ \distri_{i-1}(a) \cdot \distri_j(b) \,\cdot\, } \expQ_{k,i-2} \cdot \expQ_{i,j-1} \cdot e^{-\frac{\xi({\color{red}a, b})}{RT}} $
                \EndFor
            	}
            \EndFor
            }
        \EndFor
    \EndFor
    \State \Return $\expQ$ \Comment{Expected partition function $\expQ(\distri)=\expQ_{1,n}$}
\EndFunction
\end{algorithmic}
\end{minipage} 

\vspace{10pt}

\caption{Pseudocode for computing the classical partition function $Q(\vecx)$ for a single sequence \vecx (top, from LinearPartition \cite{zhang+:2020linearpartition})
and our expected partition function $\expQ(\distri)$ for a distribution~\distri over sequences (bottom), using the Nussinov-Jacobson energy model for simplicity of presentation. 
The former can be viewed as a special case of the latter 
when \distri is a one-hot distribution.
The changes in the latter are colored in red.
For the Turner energy model, 
besides different nonterminals as in LinearFold \cite{huang+:2019} and LinearPartition (e.g., hairpin candidates $H_{i,j}$, pairs $P_{i,j}$, etc.), 
we also need to extend $P_{i,j}$ to $P_{i,j,t}$ where 
$t\in \allpairtypes$ is the pair type of $(i, j)$, following LinearDesign \cite{zhang+:2023}.
\label{fig:pseudocode}}
\end{figure*}

Having established a distribution over RNA sequences, it is intuitive to extend this concept to define a distribution over partition functions. The \textit{partition function} $Q(\vecx)$ for a specific RNA sequence $\vecx$ is defined as:
\begin{align*}
     \quad & Q(\vecx) = \sum_{\vecy \in \GEN(\vecx)} e^{- \Delta G (\vecx, \vecy)/RT},
\end{align*}
where $\GEN(\vecx)$ represents the set of all feasible folding structures for the RNA sequence $\vecx$, and $\Delta G(\vecx, \vecy)$ is an established energy model evaluating the free energy change of mRNA $\vecx$ and structure $\vecy$. Given a distribution over RNA sequences, symbolized as $\distri$, we can derive the \textbf{expected partition function}  \(\expQ(\distri)\) over the distribution of RNA sequences as:
\begin{align*}
     \expQ (\distri) = \E_{\vecx \sim \distri(\cdot)} [Q(\vecx)] 
     &= \sum_{\vecx \in \nucset^n} \distri(\vecx)Q(\vecx)  \\
     &= \sum_{\vecx \in \nucset^n} \distri(\vecx) \Big (\sum_{\vecy \in \GEN(\vecx)}  e^{- \Delta G (\vecx, \vecy)/RT} \Big ),
\end{align*}
which represents the expected value of the partition function across all possible RNA sequences considering their respective distributions.

Fig.~\ref{fig:pseudocode} shows the pseudocode for computing the expected partition function $\expQ(\distri)$ via dynamic programming, which is a simple extension of the original \namecite{mccaskill:1990} algorithm for $Q(\vecx)$.
For simplicity of presentation, we use the Nussinov-Jacobson energy model for the pseudocode.
For the Turner energy model \cite{turner+:2009}, 
we need some non-trivial changes.
First, we need different nonterminals as in LinearFold \cite{huang+:2019} and LinearPartition (e.g., hairpin candidates $H_{i,j}$, pairs $P_{i,j}$, and multiloop candidates
$M^1_{i,j}$, $M^2_{i,j}$, $M_{i,j}$).
But more importantly, 
we also need to extend $P_{i,j}$ to $P_{i,j,t}$ where 
$t\in \allpairtypes$ is the pair type of $(i, j)$, following the implementation of lattice-parsing in LinearDesign \cite{zhang+:2023}.
This is because, for example when we extend $P_{i,j}$ to $P_{i-1,j+1}$ to form a stack, we need to know the identity of the $(i,j)$ pair and the $(i-1,j+1)$ pair, which are no longer deterministic given a distribution of sequences.
We also need this information for terminal mismatches. 

%% file: mRNA.tex


\begin{figure*}[b]
    \centering
    \begin{tabular}{m{.4\textwidth} m{.625\textwidth}} 
        \includegraphics[width=.75\linewidth]{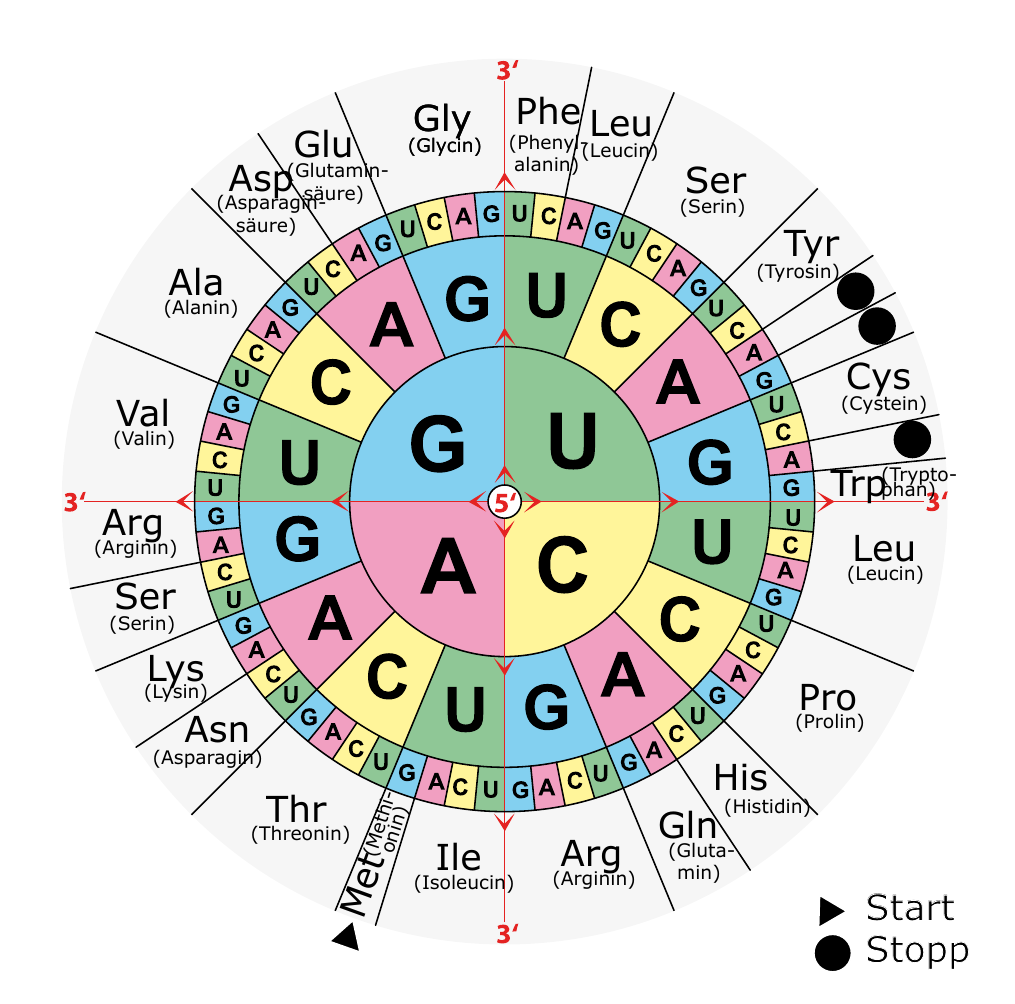} &
         \vspace{-0.1 cm}  \hspace{-1.5 cm}
         \includegraphics[width=\linewidth]{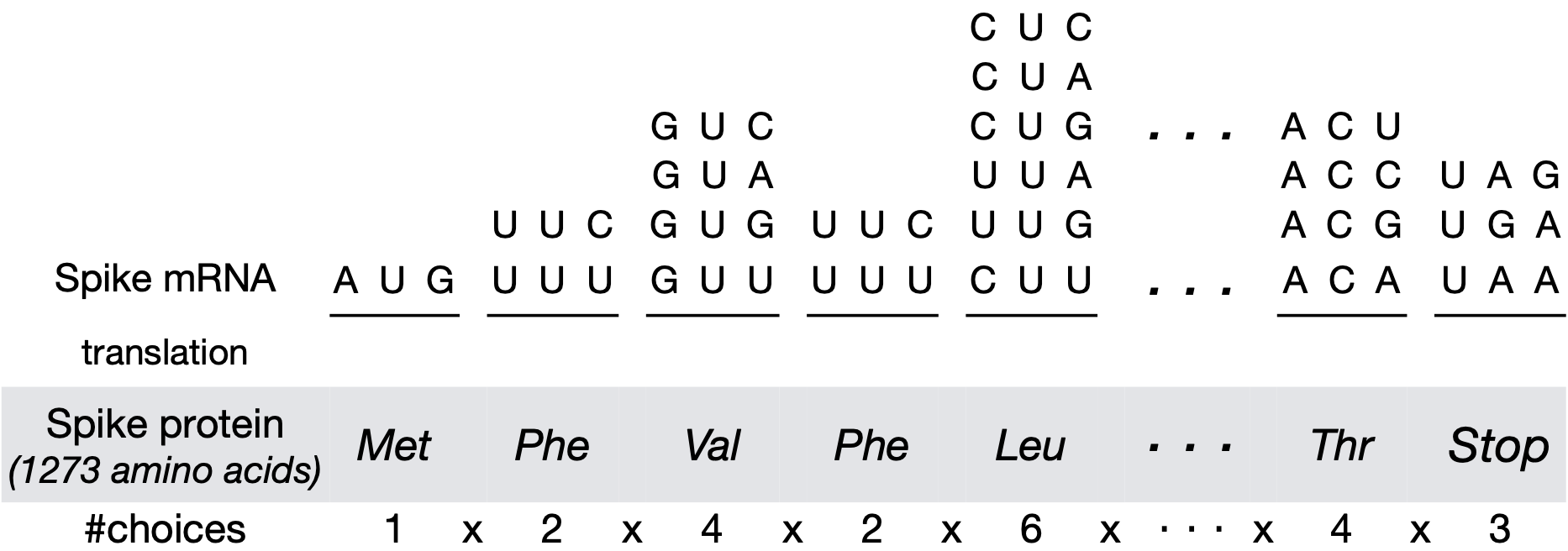} \\
    \end{tabular}
    \caption{Comprehensive Overview of mRNA Translation and Protein Synthesis. On the left, the codon wheel \cite{wikipedia2023codonwh} illustrates the correlation between mRNA codons and their respective amino acids, with the 64 possible codon combinations made from the four nucleotides: adenine (\nucA), cytosine (\nucC), guanine (\nucG), and uracil (\nucU). On the right, an example of mRNA sequence translation is depicted, showcasing the process that converts a spike protein mRNA sequence into the corresponding amino acid chain. The total choices of valid mRNA sequences are determined by the product of the number of valid codons for each amino acid at all positions.}
    \label{fig:codon_table}
\end{figure*}

In mRNA design problems, given a target protein sequence $\vecp = (p_1, p_2, ..., p_n)$ where $p_i$ denotes the $i$-th amino acid residue, we aim to find a mRNA sequence $\vecx = (x_1, x_2, .., x_{3n})$ where $x_i$ denotes the $i$-th nucleotide,  which can be translated to the target protein. During translation, every 3 nucleotides constitute a codon, and each codon specifies a particular amino acid (Figure~\ref{fig:codon_table}). 

It is not hard to see that, for a given target protein, there are exponentially many mRNA sequences that can be translated to the same protein, since an amino acid could  potentially be translated from multiple codons. As a result, the number of possible mRNA sequence candidates is the product of the kinds of condons that each amino acid in the given protein sequence can be translated from.

We also hope the designed mRNA sequence $\vecx$ satisfies some desired properties (e.g., low minimum free energy) that can be evaluated by the existing computational models. For example, LinearDesign~\citep{zhang+:2023} was devised to find the optimal mRNA sequence that has the lowest \textit{minimum folding free energy change} (MFE) among all feasible mRNA sequences. To write down the formal definition of this optimization problem, we first define a function $\text{MFE}(\vecx)$ representing the minimum folding free energy change of mRNA $\textbf{x}$ as:

\vspace{-10pt}
\begin{align*}
    \text{MFE}(\vecx) = \min_{\vecy \in \GEN(\vecx)} \Delta G (\vecx, \vecy),
\end{align*}
where $\GEN(\vecx)$ is the set of all feasible folding structures of mRNA $\vecx$, and $\Delta G(\vecx, \vecy)$ is a fixed energy model evaluating the free energy change of mRNA $\vecx$ and structure $\vecy$. This funtion is also known as the RNA folding problem \citep{huang+:2019, lorenz+:2011, do+:2006}, where we want to find the most stable structure for the input RNA $\vecx$. Based on this function, we can express the optimization problem as:

\vspace{-10pt}
\begin{align*}
    \min\limits_{\vecx} \quad & 
    \text{MFE}(\vecx) \\
    \textrm{s.t.} \quad &\vecx \in \mathcal{X}(\mathbf{p})
\end{align*}
where $\mathcal{X}(\mathbf{p})$ is the set of all feasible mRNA sequences encoding protein $\mathbf{p}$. 
 LinearDesign solved this optimization problem in  polynomial time via a dynamic programming which utilizes the local decomposable property of $\Delta G(\vecx, \vecy)$.

In this work, we investigate a modified version of the above objective. Instead of finding the mRNA sequence with the lowest $\text{MFE}(\vecx)$, we aim to find the mRNA with the largest \textit{partition function} $Q(\vecx)$. 
Instead of optimizing the single most stable structure, we want to optimize the stability of all potential folding structures simultaneously, and end up with a much smoother landscape. Formally, we can express the new optimization problem as:
\begin{equation} \label{pro:Q_x}
    \begin{split}
    \min\limits_{\vecx} \quad & - Q(\vecx) \\
    \textrm{s.t.} \quad & \vecx \in \mathcal{X}(\mathbf{p})
    \end{split}
\end{equation}

By using this new objective, we can find mRNA sequences with multiple low free energy change conformations, which could potentially lead to more stable designed mRNAs. It turns out that this new optimization formulation is NP-hard, and the orignal algorithm from LinearDesign cannot be extended to efficiently solve this problem.

Given the complexity of directly optimizing \(Q(\vecx)\), an alternative approach is to approximate this optimization problem by focusing on the \textit{expected partition function} \(\expQ(\distri)\). We propose to optimize \(\expQ(\distri)\), which reflects the average stability over a distribution of RNA sequences. This approach simplifies the problem by considering the average behavior rather than the exact configurations of each RNA sequence. The new optimization problem can then be reformulated as:
\begin{equation} \label{pro:expQ}
    \begin{split}
    \min\limits_{\distri} \quad & - \expQ(\distri) \\
    \textrm{s.t.} \quad & \distri \text{ is a valid distribution over } \nucset^n
    \end{split}
\end{equation}

In the new formulation, \(\expQ(\distri)\) can be efficiently computed via dynamic programming, similar to LinearDesign. This enables the use of gradient-based optimization methods to optimize it. 

Building on this foundation, we introduce the \textit{ensemble free energy}, \(\ensenergy(\vecx)\), for an individual RNA sequence \(\vecx\). This metric provides a more holistic view of stability by considering contributions from all competing structures, as opposed to just the Minimum Free Energy (MFE). It is defined as:
\[
\ensenergy(\vecx) = -RT \log Q(\vecx),
\]
where \(R\) is the universal gas constant, and \(T\) is the absolute temperature.

Extending this concept to our previously defined expected partition function, we introduce the \textit{generalized ensemble free energy} , \(\genEnsenergy(\distri)\). This metric captures the ensemble free energy across the expected distribution of partition functions over all possible RNA sequences. It is formulated as:
\[
\genEFE(\distri) = -RT \log \expQ(\distri),
\]
where \(\expQ(\distri)\) represents the expected partition function over the joint distribution of all positions.

Thus, the generalized ensemble free energy offers a measure of the expected stability of an ensemble of RNA sequences, taking into account not only the probabilistic distribution of the sequences but also their potential folding structures.

\input{dfa}

Following \citet{zhang+:2023}, we approach the mRNA design problem through lattice parsing. In this model, the RNA sequence is represented as a path through a lattice structure, where each node corresponds to a state in the RNA sequence construction process. To define a distribution over this lattice, we consider not only the per-position distribution of nucleotides but also the transitions between states in the lattice. Each node \(q\) in the lattice is associated with a distribution over possible outgoing transitions (edges), which represent the addition of nucleotides at that position.

For each node \(q\) in the lattice, define the distribution:
\[
\distri_{q} : \nucset(q) \mapsto [0,1], \text{s.t.~} \sum_{x \in \nucset(q)} \distri_{q} (x) = 1
\]
where \(\nucset(q)\) is the set of nucleotides permissible at node \(q\), determined by the function \(\delta(q, a)\), which defines the valid transitions from node \(q\) with nucleotide \(a\).

The overall distribution of mRNA sequences in this lattice, denoted as \(\distri : \nucset^n \mapsto [0,1]\), is thus a product of distributions at each node along the path:
\[
\distri(\vecx) = \prod_i \distri_{q_i}(x_i), \text{ given $q_0 = (0,0)$ and $q_{i+1} = \delta(q_i, x_i)$},
\]
where  $q_0 = (0,0)$ is the starting node, and  $q_{i+1} = \delta(q_i, x_i) $  is the next node reached by adding nucleotide  $x_i$.  For a more tangible understanding, a specific example is illustrated in Fig.~\ref{fig:dfa}.

%% file: dfa.tex

\colorlet{StartNode}{purple!30}
\colorlet{NodeGray}{gray!20}
\colorlet{EndNode}{purple!30}

\begin{figure*}
\centering
\begin{tikzpicture}[->,>=stealth',shorten >=1pt,auto,semithick]
\node[state, initial, initial text=, inner sep=-10pt, fill=StartNode] (n00) {\fontsize{12.5}{0}\selectfont 0,0};
\node[below=2.2cm of n00] (n0b) {$\mid$}; 

\node[state, right=.9cm of n00, inner sep=-10pt, fill=NodeGray] (n10) {\fontsize{12.5}{0}\selectfont 1,0};
\node[state, right=.9cm of n10, inner sep=-10pt, fill=NodeGray] (n20) {\fontsize{12.5}{0}\selectfont 2,0};
\node[state, right=.9cm of n20, inner sep=-10pt, fill=NodeGray] (n30) {\fontsize{12.5}{0}\selectfont 3,0};

\node[below=2.2cm of n30] (n3b) {$\mid$}; 
\node[below=2.6cm of n30] (n3c) {$\circ$}; 

\node[state, right=.9cm of n30, inner sep=-10pt, fill=NodeGray] (n40) {\fontsize{12.5}{0}\selectfont 4,0};
\node[state, below=1.2cm of n40, inner sep=-10pt, fill=NodeGray] (n41) {\fontsize{12.5}{0}\selectfont 4,1};
\node[state, right=.9cm of n40, inner sep=-10pt, fill=NodeGray] (n50) {\fontsize{12.5}{0}\selectfont 5,0};
\node[state, below=1.2cm of n50, inner sep=-10pt, fill=NodeGray] (n51) {\fontsize{12.5}{0}\selectfont 5,1};
\node[state, right=.9cm of n50, inner sep=-10pt, fill=NodeGray] (n60) {\fontsize{12.5}{0}\selectfont 6,0};

\node[below=2.2cm of n60] (n6b) {$\mid$}; 
\node[below=2.6cm of n60] (n36) {$\circ$}; 

\node[state, right=.9cm of n60, inner sep=-10pt, fill=NodeGray] (n70) {\fontsize{12.5}{0}\selectfont 7,0};
\node[state, right=.9cm of n70, inner sep=-10pt, fill=NodeGray] (n80) {\fontsize{12.5}{0}\selectfont 8,0};
\node[state, below=1.2cm of n80, inner sep=-10pt, fill=NodeGray] (n81) {\fontsize{12.5}{0}\selectfont 8,1};
\node[state, accepting, right=.9cm of n80, inner sep=-10pt, fill=EndNode] (n90) {\fontsize{12.5}{0}\selectfont 9,0};

\node[below=2.2cm of n90] (n9b) {$\mid$}; 

\draw (n0b) edge[below,<->,dashed] node{$D(\text{methionine})$} (n3b);
\draw (n3b) edge[below,<->,dashed] node{$D(\text{leucine})$} (n6b);
\draw (n6b) edge[below,<->,dashed] node{$D(\textsc{stop})$} (n9b);

\draw (n00) edge[above] node[gray]{A:1.0} (n10)
(n10) edge[above] node[gray]{U:1.0} (n20)
(n20) edge[above] node[gray]{G:1.0} (n30);

\draw (n30) edge[above] node[red]{U:0.5} (n40)
(n30) edge[above] node[red, yshift=4pt, xshift=3pt]{C:0.5} (n41)
(n40) edge[above] node[gray]{U:1.0} (n50)
(n41) edge[above] node[gray]{U:1.0} (n51)
(n50) edge[above, bend left=40] node[blue]{A:0.3} (n60)
(n50) edge[above] node[blue]{G:0.7} (n60)
(n51) edge[below, bend left=20] node[violet, yshift=4pt, xshift=-13pt]{A:0.1} (n60)
(n51) edge[below, bend right=10] node[violet, yshift=0pt, xshift=-7pt]{C:0.4} (n60)
(n51) edge[below, bend right=40] node[violet, yshift=-3pt, xshift=-12pt]{G:0.2} (n60)
(n51) edge[below, bend right=70] node[violet, yshift=-8pt, xshift=-2pt]{U:0.3} (n60);
\draw (n60) edge[above] node[gray]{U:1.0} (n70)
(n70) edge[above] node[orange]{A:0.4} (n80)
(n70) edge[above] node[orange, yshift=4pt, xshift=3pt]{G:0.6} (n81)
(n80) edge[above, bend left=40] node[cyan]{A:0.1} (n90)
(n80) edge[above] node[cyan]{G:0.9} (n90)
(n81) edge[below] node[gray, yshift=12pt, xshift=-8pt]{A:1.0} (n90);
\end{tikzpicture}

\caption{A deterministic finite automaton (DFA) representing mRNA codon sequences, with a probability distribution assigned to the transitions. Each node corresponds to a state in the DFA, with outgoing edges indicating permissible nucleotides and their associated probabilities. Nodes with a single transition, such as $(0,0), (1,0), (2,0), (4,0), (4,1), (6,0), (8,1)$, inherently have a probability of 1. For nodes with multiple outgoing edges (e.g., nodes $\textcolor{red}{(3,0)}, \textcolor{blue}{(5,0)}, \textcolor{violet}{(5,1)}, \textcolor{orange}{(7,0)}, \textcolor{cyan}{(8,0)}$), the probabilities of the outgoing edges add up to 1, ensuring a complete distribution.  Edges are color-coded to illustrate these probability groups; edges sharing the same color sum to one, denoting complete probability distributions for transitions from their respective nodes. }
\label{fig:dfa}
\end{figure*}

%% file: optimization.tex

In the mRNA design problem variant we introduced above, the goal is to optimize an objective function over all valid probability distributions \distri for RNA sequences in the design space:

\begin{equation}
\label{prob:objective}
\min_{\distri: \mathcal{X} \mapsto [0,1]} \ell(\distri)
\end{equation}
where we strive to identify the probability distribution $\distri$ that yields the minimum value for the objective function $\ell(\distri)$.

We define the parameters of the probability distribution $\distri$  by expressing it as the product of individual probabilities, thus:

\[
\distri_{\vectheta} (\vecx) = \prod_i \theta_{q_i, x_i}
\]

Here, $\theta_{q_i,x_i}$ stands for the probability of selecting the nucleotide $x_i$ at the node $q_i$ within a lattice, and \vectheta encompasses all such probabilities for each nucleotide at each node.

The optimization problem can then be formulated as follows:

\begin{equation}
\begin{aligned}\label{pro:cons}
    \min_{\boldsymbol{\theta}} \quad & 
   \ell(\distri_{\vectheta}) \\
    \textrm{subject to} \quad & \forall q, \sum_{x\in \nucset(q)} \theta_{q, x} =1
 \end{aligned}
\end{equation}

In the equation above, we aim to minimize the objective function $\ell(\distri_{\vectheta})$ subject to the constraint that for every node $q$, the sum of the probabilities of each possible nucleotide is equal to 1. This constraint ensures that each set of probabilities $\theta_{q_i}$ forms a valid multinomial distribution over the nucleotides.

Projected gradient descent provides a robust solution to the constrained optimization problem. In this method, we first compute the gradient of our objective function, $\ell(\distri_{\vectheta})$, with respect to the parameters \vectheta. The specific form of this gradient relies on the nature of $\ell$. Upon computing the gradient, we update our parameters by stepping in the negative gradient direction:

\[
\vectheta' = \vectheta^{(t)} - \alpha \nabla \ell(\distri_{\vectheta^{(t)}})
\]
where $\alpha$ is a learning rate, and $t$ denotes the current iteration.

Next, we need to project $\vectheta'$ back onto the set of valid multinomial distributions. This can be done by projecting the parameters $\vectheta_q$ in each node $q$ back to the probability simplex, which is equivalent to solving to following problem:
\begin{align*}
\vectheta_q^{(t+1)} = \argmin_{\vectheta_{q}} \quad & \|\vectheta_{q} - \vectheta'_{q}\|^2_2 \\
    \textrm{subject to} \quad & \sum_{x\in \nucset(q)} \theta_{q, x} =1, \quad \theta_{q, x} \ge 0, \forall x \in \nucset
\end{align*}
The equation above represents the projection of $\vectheta^{(t+1)}_{q}$ onto the probability simplex, where $\|\cdot\|^2_2$ denotes the Euclidean norm. The constraints ensure that $\theta_{q}$ is a valid probability distribution over $\nucset$. The solution to this problem can be efficiently computed using algorithms for simplex projections.

This process of gradient update and projection is repeated until the changes in $\vectheta$ or the value of the objective function $\ell(\distri_{\vectheta})$ becomes sufficiently small, which indicates that the solution has converged.

Algorithm~\ref{alg:PGD} below presents a summary of the projected gradient descent method for the constrained optimization problem:

\begin{algorithm}
\caption{Projected Gradient Descent}
\label{alg:PGD}
\begin{algorithmic}
\State Initialize $\vectheta^{(0)}$ such that each $\theta_{q_i,x_i}^{(0)}$ is a valid probability distribution
\While{not converged}
    \State Compute gradient $\nabla \ell(\distri_{\vectheta^{(t)}})$
    \State Update $\vectheta^{(t+1)} = \vectheta^{(t)} - \alpha \nabla \ell(\distri_{\vectheta^{(t)}})$
    \State Project $\vectheta^{(t+1)}$ onto the probability simplex:
    \Indent 
    \For{each $q$}
        \State solve the projection problem to update $\theta_{q}^{(t+1)}$
    \EndFor
    \EndIndent
    \State Update $t = t + 1$
\EndWhile
\State \Return Solution $\vectheta^*$ to the optimization problem
\end{algorithmic}
\end{algorithm}

%% file: results-mRNA.tex

In this section, we evaluate our proposed approach through a series of experiments designed to assess its effectiveness and characteristics. We conducted a random sampling of 20 natural protein sequences from the UniProt database \cite{UniProt}. These sequences exhibit a diverse range of lengths, varying from 50 to 350 amino acids, which allows for a comprehensive analysis across different protein sizes. The specific sequence IDs, along with their corresponding length statistics, are detailed in Table~\ref{tb:results}. 

\paragraph{\textbf{Implementation Details}} 
We adapted and modified the mRNA design lattice parsing framework originally developed by \citet{zhang+:2023}. Our extensions primarily focus on the computation of the expected partition function, adopting a methodology akin to that used in LinearPartition~\cite{zhang+:2020linearpartition}. Additionally, we also incorporated the beam pruning heuristic to accelerate this computation. Regarding the calculation of gradients, we leveraged the properties of the inside-outside algorithm, which allows us to derive the gradients directly by combining the results from both the inside and outside phases of the algorithm, as detailed in \citet{eisner:2016}.

Throughout the development and experimental phases, we observed that the projected gradient method, which primarily considers local statistics of the optimization problem, is significantly influenced by the initial distribution, denoted as $\vectheta^{(0)}$, in Algorithm~\ref{alg:PGD}. The choice of $\vectheta^{(0)}$ has a substantial impact on the final solution attained by our program. To address this, we implemented the following strategies:

\begin{itemize}
\item \textbf{Initializing Close to MFE Solution:} Rather than starting from a random distribution $\distri_\text{rand}$, which often leads to suboptimal solutions especially for longer proteins, we propose using the MFE solution $\distri_\text{MFE}$ (a distribution contains only the MFE mRNA sequence) from LinearDesign~\cite{zhang+:2023} as the initial point. However, directly initializing the initial distribution $\distri_{\vectheta^{(0)}}$ with the MFE solution may lead to the algorithm becoming trapped in the MFE solution. To mitigate this, we introduce a \textit{soft-MFE} initialization strategy:
\[
\distri_\text{soft-MFE}(\varepsilon) = \varepsilon \distri_\text{rand} + (1 - \varepsilon) \distri_\text{MFE},
\]
blending the random distribution $\distri_\text{rand}$ with the MFE solution $\distri_\text{MFE}$, controlled by the parameter $\varepsilon \in [0, 1]$. Setting $\varepsilon$ to 0 reduces $\distri_\text{soft-MFE}$ to MFE initialization $\distri_\text{MFE}$, while setting $\varepsilon$ to 1 transforms it into random initialization $\distri_\text{rand}$.

\item \textbf{Parallelization with Multiple Runs:} Another strategy is to experiment with various initializations of $\vectheta^{(0)}$, executing the gradient descent process in parallel. The optimal solution can then be selected from the outcomes of these concurrent runs.
\end{itemize}

\paragraph{\textbf{Baseline Method}}
For comparative purposes, we implemented a \textit{random walk} baseline. In this approach, we continuously refine the best mRNA sequence we have identified by randomly altering one codon at a time. Initially, the best mRNA sequence is set to the MFE solution from LinearDesign. At each step, we randomly select an amino acid and substitute its corresponding mRNA segment with a random codon. The objective value of the new mRNA sequence is then evaluated; if it shows an improvement over the current best sequence, it is recorded as the new optimal sequence. This process is repeated until a predetermined number of steps is reached.

\paragraph{\textbf{Main Results}}

\begin{table*}[!h]
\caption{ Comparative Results of Baseline and Proposed Methods across Protein Sequences. This table encapsulates the best results of 20 independent runs per protein sequence for both the random walk baseline and our proposed soft-MFE method with $\varepsilon=0.5$, beam size {\it b}=100. The evaluation metric is the Ensemble Free Energy (\EFE). For each protein, identified by UniProt ID and length (in amino acids), we report the \EFE of the best-performing mRNA sequence as determined by LinearPartition. The table includes both the absolute \EFE and the change in \EFE (\DeltaEFE) compared to the MFE solution, as well as the percentage of changed codons.}
  \label{tb:results}
  \centering
  \begin{small}
  \begin{tabular}{lcc||rrr||rrr||rrr}
    \toprule
     \multirow{2}{*}{UniProt} & \multirow{3}{*}{Length} & \multirow{2}{*}{LinearDesign} 
     & \multicolumn{3}{c||}{Random Walk} 
     & \multicolumn{3}{c||}{soft-MFE $\varepsilon$=0.5, {\it b}=100 } 
     & \multicolumn{3}{c}{soft-MFE $\varepsilon$=0.5, {\it b}=200 } \\
     \cmidrule(r){4-6} \cmidrule(r){7-9} \cmidrule(r){10-12}
      \quad ID & & MFE solution 
          & \multirow{2}{*}{\EFE} & \multirow{2}{*}{\DeltaEFE} & {\% codon } 
          & \multirow{2}{*}{\EFE} & \multirow{2}{*}{\DeltaEFE} & {\% codon } 
          & \multirow{2}{*}{\EFE} & \multirow{2}{*}{\DeltaEFE} & {\% codon} \\
     \cmidrule(r){2-2} \cmidrule(r){3-3}  
          & protein / mRNA & \EFE & & & changed  & & & changed  & & &  changed \\
    \midrule
{\tt Q13794} & 54  \aas \,/ 162 \nts& -113.39 & -113.58 & {\bf -0.19} & 3.7 & -113.58 & {\bf -0.19} & 3.7  & -113.58 & {\bf -0.19} & 3.7 \\
{\tt Q9UI25} & 63  \aas \,/ 189 \nts& -126.07 & -126.23 & {\bf -0.16} & 7.9 & -126.23 & {\bf -0.16} & 7.9  & -126.23 & {\bf -0.16} & 7.9 \\
{\tt Q9BZL1} & 73  \aas \,/ 219 \nts& -114.87 & -115.12 & {    -0.25} & 6.9 & -115.25 & {\bf -0.38} & 5.5  & -115.25 & {\bf -0.38} & 6.9 \\
{\tt P60468} & 96  \aas \,/ 288 \nts& -234.36 & -234.50 & {    -0.14} & 2.1 & -234.74 & {\bf -0.38} & 4.2  & -234.74 & {\bf -0.38} & 4.2 \\
{\tt Q9NWD9} & 120 \aas \,/ 360 \nts& -226.36 & -226.36 & {     0.00} & 0.8 & -227.03 & {\bf -0.67} & 4.2  & -227.03 & {\bf -0.67} & 4.2 \\
{\tt P14555} & 144 \aas \,/ 432 \nts& -275.06 & -275.06 & {     0.00} & 0.0 & -275.47 & {\bf -0.41} & 3.5  & -275.47 & {\bf -0.41} & 3.5 \\
{\tt Q8N111} & 149 \aas \,/ 447 \nts& -335.89 & -335.94 & {    -0.05} & 0.7 & -336.40 & {\bf -0.51} & 5.4  & -336.40 & {\bf -0.51} & 4.7 \\
{\tt P63125} & 156 \aas \,/ 468 \nts& -299.18 & -299.60 & {\bf -0.42} & 3.2 & -299.59 & {    -0.41} & 5.1  & -299.59 & {    -0.41} & 5.1 \\
{\tt Q6XD76} & 172 \aas \,/ 516 \nts& -427.19 & -427.50 & {    -0.31} & 1.2 & -427.82 & {\bf -0.63} & 2.9  & -427.82 & {\bf -0.63} & 1.7 \\
{\tt P0DMU9} & 189 \aas \,/ 567 \nts& -361.48 & -362.10 & {    -0.62} & 2.7 & -362.12 & {\bf -0.64} & 3.7  & -362.12 & {\bf -0.64} & 3.7 \\
{\tt P0DPF6} & 209 \aas \,/ 627 \nts& -545.55 & -545.97 & {    -0.42} & 1.0 & -546.07 & {\bf -0.52} & 1.9  & -546.07 & {\bf -0.52} & 1.9 \\
{\tt Q9HD15} & 224 \aas \,/ 672 \nts& -532.98 & -533.19 & {    -0.21} & 2.2 & -522.84 & {    10.14} & 18.8 & -533.30 & {\bf -0.32} & 3.6 \\
{\tt Q6T310} & 242 \aas \,/ 726 \nts& -504.29 & -504.53 & {    -0.24} & 1.2 & -504.82 & {\bf -0.53} & 3.7  & -504.82 & {\bf -0.53} & 3.3 \\
{\tt Q9BRP0} & 275 \aas \,/ 825 \nts& -586.50 & -586.84 & {    -0.34} & 0.7 & -584.39 & {     2.11} & 29.5 & -587.49 & {\bf -0.99} & 3.6 \\
{\tt P56178} & 289 \aas \,/ 867 \nts& -606.58 & -607.04 & {    -0.46} & 0.4 & -607.50 & {\bf -0.92} & 2.8  & -607.50 & {\bf -0.92} & 2.8 \\
{\tt Q8NH87} & 305 \aas \,/ 915 \nts& -572.37 & -572.83 & {    -0.46} & 1.0 & -573.43 & {\bf -1.06} & 3.6  & -573.43 & {\bf -1.06} & 3.9 \\
{\tt Q8NGU1} & 314 \aas \,/ 942 \nts& -612.55 & -612.67 & {    -0.12} & 1.3 & -610.36 & {     2.19} & 8.6  & -613.42 & {\bf -0.87} & 5.1 \\
{\tt Q8NGC9} & 324 \aas \,/ 972 \nts& -582.39 & -582.86 & {    -0.47} & 1.9 & -578.51 & {     3.88} & 23.5 & -583.43 & {\bf -1.04} & 4.3 \\
{\tt Q99729} & 332 \aas \,/ 996 \nts& -667.21 & -667.74 & {    -0.53} & 0.6 & -668.21 & {\bf -1.00} & 3.0  & -668.21 & {\bf -1.00} & 3.0 \\
{\tt Q9P2M1} & 347 \aas \,/ 1041\nts& -663.54 & -664.15 & {    -0.61} & 0.9 & -664.95 & {    -1.41} & 5.5  & -665.07 & {\bf -1.53} & 5.5 \\

    \bottomrule
 \end{tabular}
 \end{small}
\end{table*}

In our principal experiments, we conducted 20 independent runs for each protein sequence using both the baseline method and our proposed method. Each run was initialized with a distinct random seed, ensuring consistency across different methods and settings. The final reported result, \EFE, for each method and setting was determined by selecting the best mRNA sequence evaluated by LinearPartition (using the \verb|-V -b0 -d0| options) from all the runs. For the baseline method, we implemented a 100-step random walk for each run. In contrast, our method employed a 30-step projected gradient descent, as we consistently observed model convergence within this number of iterations for proteins of varying lengths. Regarding the initialization parameter $\varepsilon$ in our method, we consistently used $\varepsilon = 0.5$, as empirical testing revealed this to be the most effective setting across different proteins.

\begin{figure}[h]
\centering
	
  \begin{subfigure}[b]{0.45\textwidth}
      \resizebox{\linewidth}{!}{
\begin{tikzpicture}
\begin{axis}[
    width=14cm,
    height=6cm,
    xlabel={Protein Length (amino acids)},
    ylabel={\% Codon Changed},
    xmin=45, xmax=355,
    ymin=-0.05, ymax=9,
    xtick={50,100,150,200,250,300,350},
    ytick={0, 1, 2, 3, 4, 5, 6, 7, 8},
    legend pos=north east,
    ymajorgrids=true,
    grid style=dashed,
]

\addplot[
    color=blue,
    mark=square,
    ]
    coordinates {
    (54, 3.7) (63, 7.9) (73, 6.9) (96, 2.1) (120, 0.8)
    (144, 0.0) (149, 0.7) (156, 3.2) (172, 1.2) (189, 2.7)
    (209, 1.0) (224, 2.2) (242, 1.2) (275, 0.7) (289, 0.4)
    (305, 1.0) (314, 1.3) (324, 1.9) (332, 0.6) (347, 0.9)
    };
    \addlegendentry{Baseline}

\addplot[
    color=red,
    mark=triangle,
    ]
    coordinates {
    (54, 3.7) (63, 7.9) (73, 6.9) (96, 4.2) (120, 4.2)
    (144, 3.5) (149, 4.7) (156, 5.1) (172, 1.7) (189, 3.7)
    (209, 1.9) (224, 3.6) (242, 3.3) (275, 3.6) (289, 2.8)
    (305, 3.9) (314, 5.1) (324, 4.3) (332, 3.0) (347, 5.5)
    };
    \addlegendentry{Ours ({\it b}=200)}

\end{axis}
\end{tikzpicture}
}
      \caption{Percentage of Codon Change vs. Protein Length.}
      \label{fig:codon_vs_len}
      
  \end{subfigure}

      \begin{subfigure}[b]{0.45\textwidth}
      \resizebox{\linewidth}{!}{
\begin{tikzpicture}
\begin{axis}[
    width=14cm,
    height=6cm,
    xlabel={Protein Length (amino acids)},
    ylabel={\(\Delta\EFE\)},
    xmin=45, xmax=355,
    ymin=-1.75, ymax=0.5,
    xtick={50,100,150,200,250,300,350},
    ytick={-1.75,-1.5,-1.25,-1.0, -0.75, -0.5, -0.25,0, 0.25, 0.5},
    legend pos=north east,
    ymajorgrids=true,
    grid style=dashed,
]

\addplot[
    color=blue,
    mark=square,
    ]
    coordinates {
    (54, -0.19) (63, -0.16) (73, -0.25) (96, -0.14) (120, 0.00)
    (144, 0.00) (149, -0.05) (156, -0.42) (172, -0.31) (189, -0.62)
    (209, -0.42) (224, -0.21) (242, -0.24) (275, -0.34) (289, -0.46)
    (305, -0.46) (314, -0.12) (324, -0.47) (332, -0.53) (347, -0.61)
    };
    \addlegendentry{Baseline}

\addplot[
    color=red,
    mark=triangle,
    ]
    coordinates {
    (54, -0.19) (63, -0.16) (73, -0.38) (96, -0.38) (120, -0.67)
    (144, -0.41) (149, -0.51) (156, -0.41) (172, -0.63) (189, -0.64)
    (209, -0.52) (224, -0.32) (242, -0.53) (275, -0.99) (289, -0.92)
    (305, -1.06) (314, -0.87) (324, -1.04) (332, -1.00) (347, -1.53)
    };
    \addlegendentry{Ours ({\it b}=200)}

\end{axis}
\end{tikzpicture}
}
      \caption{Change in Ensemble Free Energy (\(\Delta\EFE\)) vs. Protein Length.}
      \label{fig:DeltaEFE_vs_len}

  \end{subfigure}
  \caption{Comparative Analysis of Codon Changes and Ensemble Free Energy Variations. These subfigures collectively display the performance of the baseline method and our proposed method (beam size of 200) in terms of codon variability and \DeltaEFE across a range of protein lengths, according to the results in Table~\ref{tb:results}.}
\end{figure}

Table~\ref{tb:results} presents the comparative results of these different methods. At a glance, the random walk method provides a robust baseline, generally yielding mRNA sequences with lower ensemble free energy (\EFE) across the tested protein sequences. However, an analysis of the percentage change in codons compared to the MFE solution reveals that the random walk baseline tends to explore sequences in close proximity to the MFE solution (typically within about 1-2\% change for longer proteins), indicating its potential limitations for longer proteins.

In contrast to the baseline, our method demonstrated the ability to identify more diverse mRNA solutions, differing more significantly from the MFE solution than those found by the baseline. This is particularly evident in the case of longer proteins, where the best solutions from our method display a 3-5\% variation in codons relative to the MFE solutions, as depicted in Fig.\ref{fig:codon_vs_len}.  Additionally, our method not only explores a broader range of mRNA sequences than the baseline but also consistently achieves higher ensemble free energy for the best-found sequences. Moreover, in scenarios with a beam size of 200, there is a discernible trend: for longer sequences, our method tends to yield solutions with greater improvements compared to the baseline method, as illustrated in Fig.~\ref{fig:DeltaEFE_vs_len}. These findings indicate the potential advantages of our method, especially for sequences of extended length.

\begin{figure}[h]
    \centering
       \includegraphics[width=\linewidth]{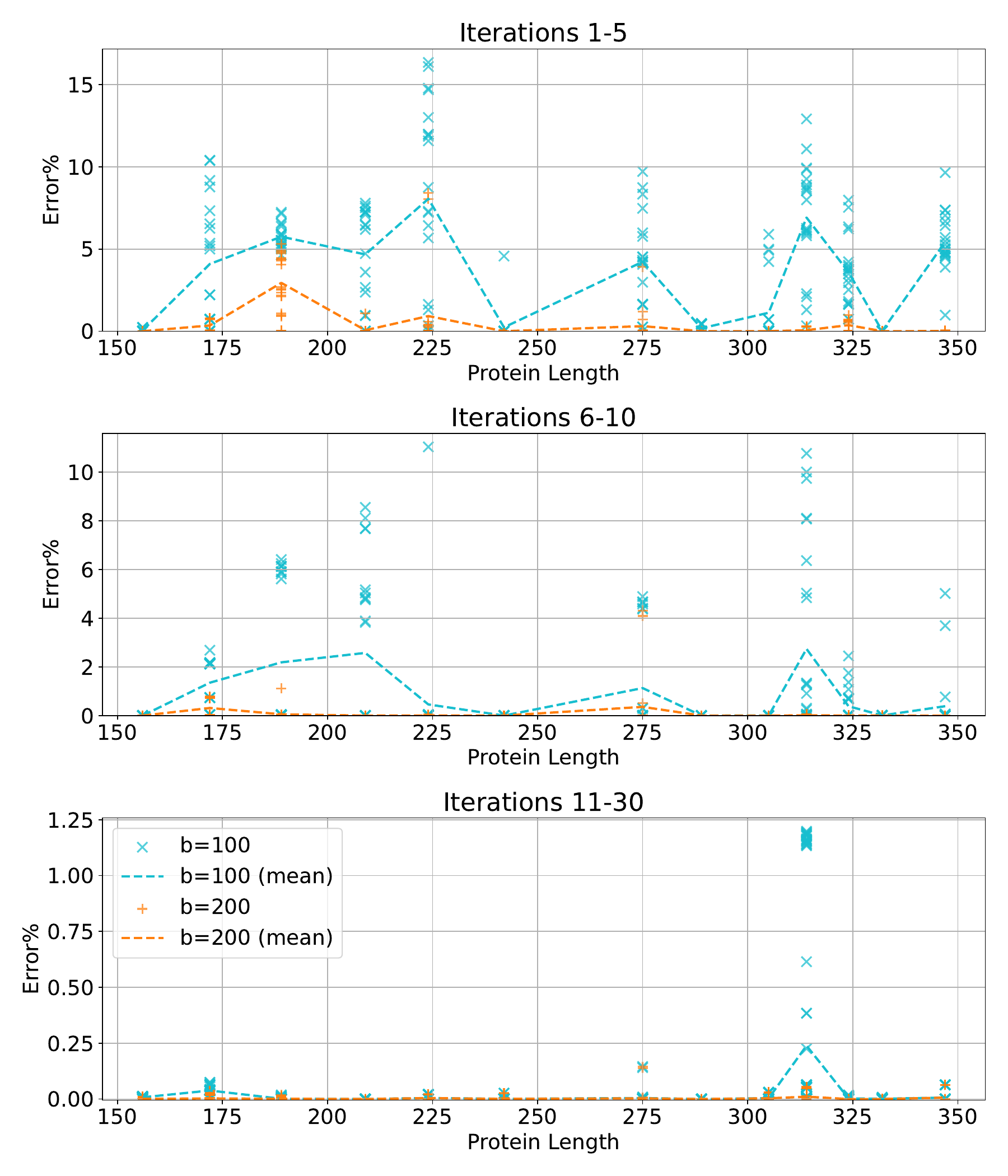} 
    \caption{Beam Pruning Search Error Relative to Protein Length. This figure illustrates the relative search error for proteins exceeding 150 amino acids, derived from 5 independent runs using varying beam sizes. Each run comprised 30 optimization steps. We categorized the iterations into three groups: 1-5, 6-10, and 11-30, to highlight the evolution of error across different optimization stages. Initially, search errors are higher, reflecting early-stage optimization challenges. Notably, the error does not escalate with increasing protein length, indicating the pruning's effectiveness and stability across varying sequence lengths.}
    \label{fig:beam_err}
\end{figure}

\paragraph{\textbf{Beam Pruning Search Error Analysis}}
In this part, we aimed to quantify the search error introduced by the beam pruning method. To achieve this, we tracked and recorded the objective values computed at various beam sizes throughout the 30-step optimization process. Our analysis focused on comparing these values with the `true' objective value, which we obtained in the absence of beam pruning. The results, illustrated in Fig.~\ref{fig:beam_err}, display the relative search error across different sequence lengths and various stages of optimization.

\begin{figure}[h]
    \begin{subfigure}[b]{0.43\textwidth}
        \includegraphics[width=1.1\textwidth]{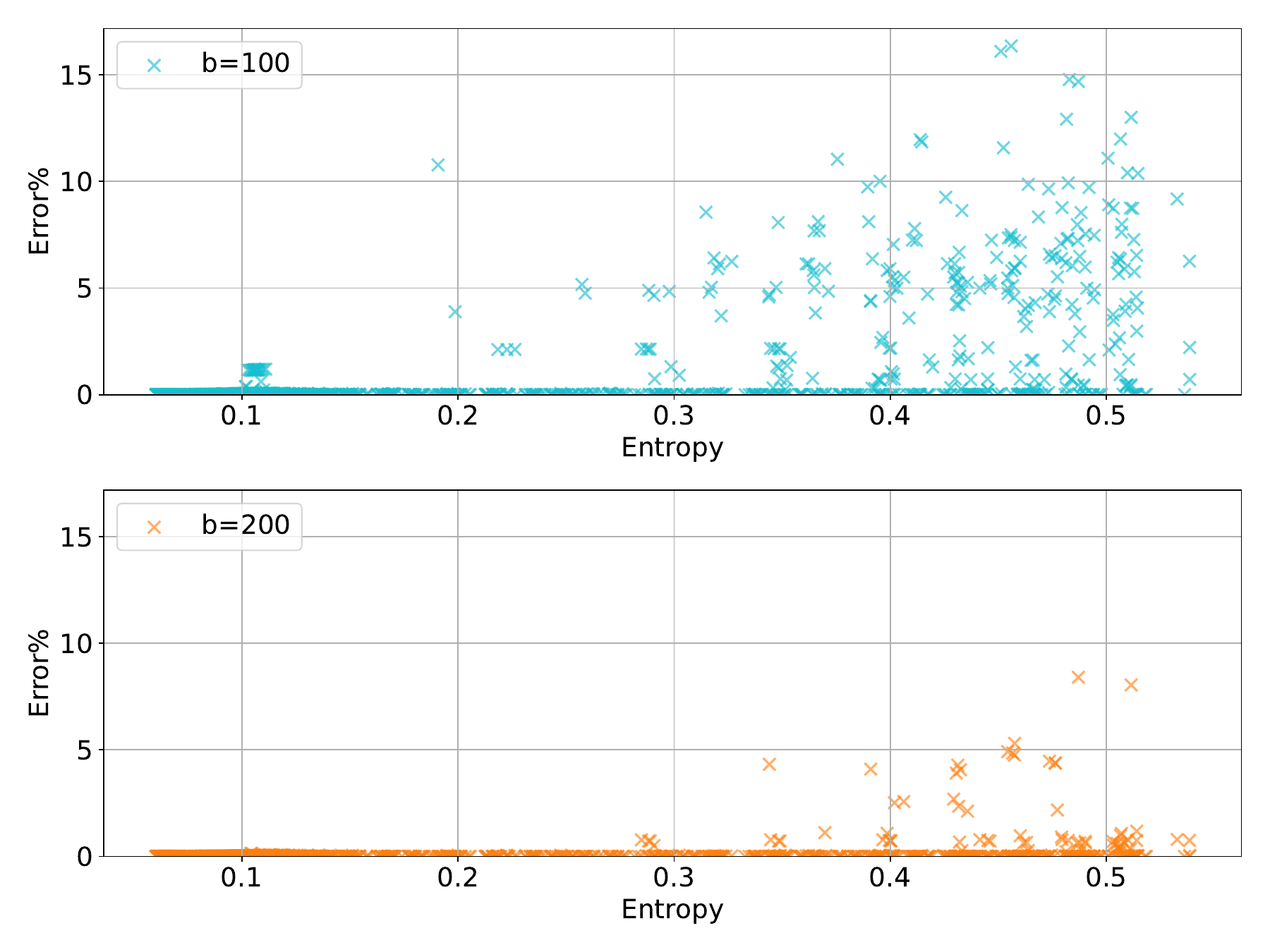}
        \caption{Search Error\% vs. Entropy}
        \label{fig:err_vs_ent}
    \end{subfigure}
   
    \begin{subfigure}[b]{0.43\textwidth}
        \includegraphics[width=1.23\textwidth]{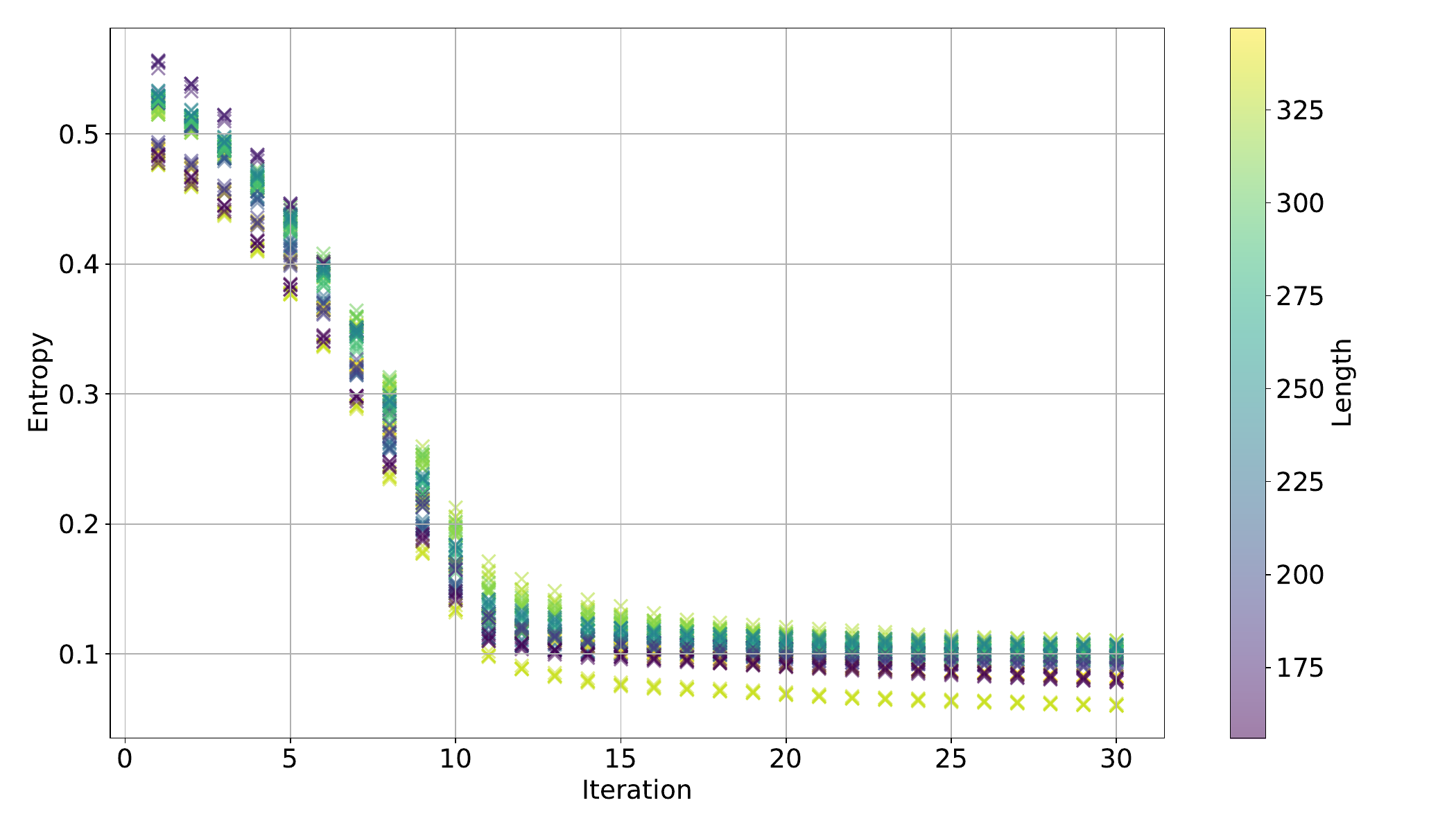}
        \caption{Entropy vs. Iteration}
        \label{fig:ent_vs_iter}
    \end{subfigure}
    \caption{The interplay between Entropy and Search Error during the Optimization Process. On the left, the correlation between search error and entropy for a fixed beam size is shown (Fig.~\ref{fig:err_vs_ent}). On the right, the progression of entropy values through different iterations is displayed (Fig.~\ref{fig:ent_vs_iter}), demonstrating a general decrease in entropy as optimization advances, irrespective of protein length.}
    \label{fig:err_ent_iter}
\end{figure}

The findings reveal that search errors tend to be more significant during the early stages of optimization when the distributions are relatively `soft', characterized by higher entropy (see Fig.~\ref{fig:ent_vs_iter}). For instance, in the initial iterations (1-5), we observed a maximum relative error of approximately $16\%$, with an average error around $5\%$. In contrast, during the later stages of the process (iterations 11-30), the maximum relative error notably decreases to below $1.25\%$, and the mean error falls to less than $0.25\%$. This trend suggests that the impact of beam pruning on search accuracy diminishes as the optimization progresses and the distributions become more `focused' or `sharpened.'

In our subsequent analysis, we delved deeper into the relationship between the entropy of distributions and the search error attributable to beam pruning during the optimization process. The data points for this analysis were consistent with those used in the previous experiments. In this context, we quantify the entropy of a distribution as the average entropy over all nodes in the mRNA DFAs. The entropy of a specific node in a DFA, for a given distribution \(\distri\), is mathematically computed as follows:

\[
\text{Entropy}(\distri, q) = - \sum_{x \in \nucset(q)} \distri_{q} (x) \log_2{\distri_{q} (x)},
\]
where \(q\) represents a node in the mRNA DFA, and \(\nucset(q)\) denotes the set of nucleotides that are permissible at node \(q\). This set is determined by the transition function \(\delta(q, a)\), which specifies the valid transitions from node \(q\) given nucleotide \(a\).

As depicted in Fig.~\ref{fig:err_vs_ent}, there is a visible trend that, for a given beam size, the relative search error increases together with the entropy of the distribution. This observation underpins the idea that higher uncertainty (entropy) in the distribution leads to greater search errors during optimization. Furthermore, Fig.~\ref{fig:ent_vs_iter} provides a clear illustration of the entropy dynamics throughout the optimization process. Notably, at the start of optimization, the entropy values are at their peak, gradually decreasing as the process progresses. This trend is consistent regardless of the protein length involved in the study.

\paragraph{\textbf{Case Study on Long Protein Sequences}}

\begin{figure*}[!ht]
    \centering
       \includegraphics[width=\linewidth]{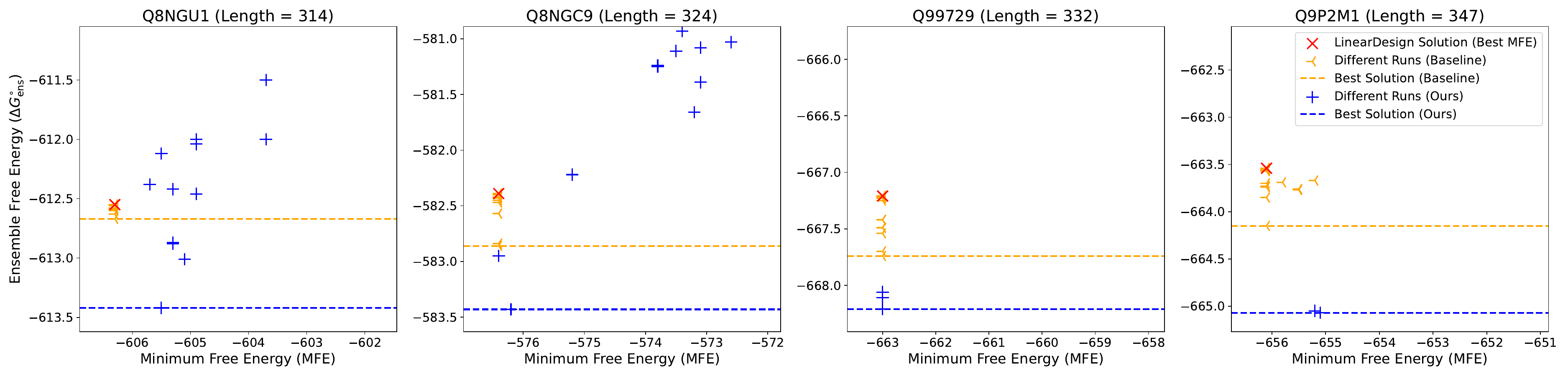} 
    \caption{Comparative 2D plots of Minimum Free Energy (MFE) and Ensemble Free Energy (EFE) values for the final solutions from 20 runs for each of the four longest protein sequences in the dataset. The focus is on the region near the MFE solution, where distinct points represent unique solutions. Some points overlap due to identical solutions found across multiple runs, while others may not be visible if they fall outside the zoomed-in area, usually indicating a higher Ensemble Free Energy compared to the MFE solution. }
    \label{fig:multi_runs}
\end{figure*}

In this case study, we focus on four of the longest proteins examined in our experiments: \texttt{Q8NGU1}, \texttt{Q8NGC9}, \texttt{Q99729}, and \texttt{Q9P2M1}. The results of all 20 runs for each sequence were evaluated. We visualize the outcomes as points on a 2D plane based on their Minimum Free Energy (MFE) and Ensemble Free Energy (EFE) values, as calculated by LinearFold and LinearPartition, respectively (see Fig.~\ref{fig:multi_runs}).

A key observation from the results is the tendency of the baseline method to yield solutions clustered around the MFE solution. This pattern is consistent with our earlier findings (Fig.~\ref{fig:codon_vs_len}), indicating the baseline's limited exploration range, primarily near the initial MFE solution. Conversely, our method demonstrates a broader search capability, generating more diverse solutions that significantly deviate from the MFE baseline.

While our approach outperforms the baseline in terms of solution diversity, it's crucial to note that it can sometimes converge to suboptimal solutions with higher EFE than the MFE solution. This outcome contrasts with the baseline method, which consistently improves or matches the MFE solution. Our method's divergence from the initial distribution during gradient descent may lead to local optima without prior knowledge of their EFE values. To address the risk of converging to suboptimal solutions, we emphasize the importance of multiple initializations and runs. By exploring various starting points, our method increases the likelihood of discovering superior solutions, particularly vital for long protein sequences.

\quad

\paragraph{\textbf{Results on the SARS-CoV-2 Spike Protein}}

\begin{table}[h]
\caption{Results of LinearDesign, Baseline, and Our Methods on the SARS-CoV-2 Spike Protein.}
  \label{tb:covid_results}
  \centering
  \resizebox{.49\textwidth}{!}{%
  \begin{small}
  \begin{tabular}{c|rrr|rrr}
    \toprule
      \multirow{2}{*}{LinearDesign} 
     & \multicolumn{3}{c|}{Random Walk} 
     & \multicolumn{3}{c}{soft-MFE $\varepsilon$=0.5, {\it b}=300 } \\
     \cmidrule(r){2-4} \cmidrule(r){5-7} 
        MFE solution 
          & \multirow{2}{*}{\EFE} & \multirow{2}{*}{\DeltaEFE} & {\% codon } 
          & \multirow{2}{*}{\EFE} & \multirow{2}{*}{\DeltaEFE} & {\% codon} \\
     \cmidrule(r){1-1}   
           \EFE & & & changed  & & & changed \\
    \midrule
  -2511.49 & -2512.31 & { -0.82} & 0.2 & -2515.84 & {\bf -4.35} & 3.8   \\

    \bottomrule
 \end{tabular}
 \end{small}
 }
\end{table}

\begin{figure}
    \centering
       \includegraphics[width=.75\linewidth]{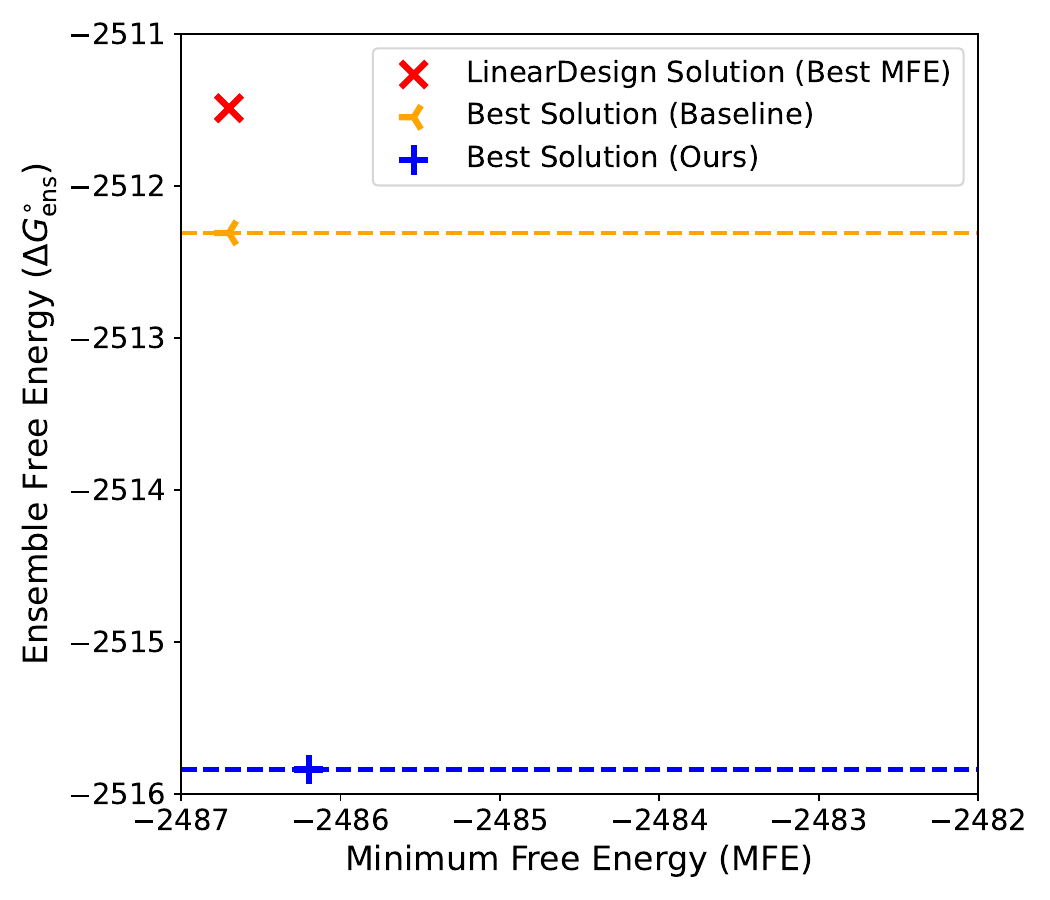} 
    \caption{Comparison of Minimum Free Energy (MFE) and Ensemble Free Energy (EFE) for the best solutions from LinearDesign, Baseline, and Our Method on the SARS-CoV-2 Spike Protein.}
    \label{fig:covid_plot}
\end{figure}

We further test our algorithm on the SARS-CoV-2 Spike Protein, comprising 1,273 amino acids, and compare it with both the Random Walk baseline and LinearDesign. As indicated in Table~\ref{tb:covid_results}, our method's best solution exhibits a $3.8\%$ codon alteration, significantly surpassing the baseline's $0.2\%$ variation from the MFE solution. This underscores our method's advantage for long sequences. Additionally, our method markedly improves the Ensemble Free Energy, as depicted in Fig.~\ref{fig:covid_plot}. While aiming for superior ensemble free energy, our solution accepts a trade-off in minimum free energy, unlike the baseline, which maintains the MFE but with lesser ensemble free energy improvement.

%% file: related.tex

The recent work by \namecite{matthies+:2023}
proposes a framework very similar to our expected partition function, which they call the ``structure-sequence partition function'',
but from a  different motivation
and for the different
task 
of non-coding (ncRNA) RNA design.
They first defined the concept of  ``sequence partition function'' as $$\expQy (\distri) = \sum_{\vecx} \distri(\vecx) e^{-\Delta G(\vecx,\vecy) / RT} =  \E_{\vecx \sim \distri(\cdot)}[e^{-\Delta G(\vecx, \vecy) / RT}] ,$$ 
and then extended it to the ``structure-sequence partition function'': 
$$\expQ (\distri) =  \sum_{\vecy}  \expQy (\distri) .$$ 

We argue that this framework is a better fit for mRNA design than ncRNA design because as \namecite{zhang+:2023} showed,
the mRNA design space can be categorized by a regular language (or finite-state automata) and the objective function (either MFE or partition function) can be formulated as a stochastic context-free grammar (SCFG).
It is well-known in formal language theory that the intersection of a (stochastic) context-free language and a regular language is another (stochastic)  context-free language, which is the essential reason why 
\namecite{zhang+:2023}'s algorithm for MFE-based mRNA design 
runs in worst-case $O(n^3)$ time.
By contrast, the design space for ncRNA design given a target structure \vecy is not a regular, but rather a context-free language, because
the paired positions in \vecy are correlated
(for any pair $(i,j)\in \vecy$, $x_i x_j$ must be pairable; it is straightforward to write a context-free grammar for the set of 
all possible \vecx for $\vecy$). 
Unfortunately, the intersection of two context-free languages are context sensitive, which rules out polynomial-time algorithms. This observation suggests that ncRNA design is fundamentally harder than mRNA design, and this framework is more promising for mRNA design.


%% file: conclusions.tex

We described a general framework of continuous optimization for RNA design problems based on the concept of expected partition function, which is a generalization of classical partition function to a distribution of sequences.
We showed how to use this framework for an important use case,
the mRNA design problem optimizing for partition function (i.e., ensemble free energy) which is much harder than the optimization for MFE.
Our results on 20 protein sequences showed that our approach can consistently improve over LinearDesign's results, 
with bigger improvements on longer sequences.


%% file: fig_LD_partition.tex

\begin{figure*}[!htb]
\hspace{-15pt}
\centering
\small
\begin{minipage}{.46\textwidth}
\begin{algorithmic}
\Function{LD-Partition}{$D, b$} \Comment{$b$ is beam size}
    \State $n \gets$ length of $D$
    \State $Q^{S} \gets$ hash() \Comment{hash table: from state $[q_i, q_j]$ to $\Qsf{S}{q_i}{q_j}$}
    \State $Q^{P} \gets$ hash() \Comment{hash table: from state $[q_i, q_j]$ to $\Qsf{P}{q_i}{q_j}$}
    \For{$j=4 \ldots n$}
      \For {$q_{j-1} \in \nodes(D, j-1)$ and $q_i$ s.t. $(q_i, q_{j-1}) \in \expQ^S$}
         	\For {$\laedgew{q_{j-1}}{b}{q_j} \in \outedges(D, q_{j-1})$} 
            \State $\Qsf{S}{q_i}{q_j} \pluseq \Qsf{S}{q_i}{q_{j-1}} \cdot e^{-\frac{\delta(b)}{RT}} $
		     \For {$\laedgew{q_{i-1}}{a}{q_i} \in \inedges(D, q_i)$} 
                       \If {$a, b$ in \allpairtypes}
                       \State $\Qsf{P}{q_{i-1}}{q_j} \pluseq \Qsf{S}{q_i}{q_{j-1}} \cdot e^{-\frac{\xi(a, b)}{RT}}$ 
                     \EndIf

		   \EndFor
	        \EndFor
	     \EndFor
      \State \textproc{BeamPrune}$(Q^S, Q^P, j, b)$ \Comment{top-$b$ among all $\Qsf{P}{q_i}{q_j}$}
      \For {$q_j \in \nodes(D, j)$}
         \For {$q_i$ such that $(q_i, q_j) \in Q^P$}
            \For {$q_k$ such that $(q_k, q_i) \in Q^S$}
                \State $\Qsf{S}{k}{j} \pluseq \Qsf{S}{k}{i} \cdot \Qsf{P}{i}{j}$ 

           \EndFor          
         \EndFor
      \EndFor
      \State \textproc{BeamPrune}$(Q^S, Q^S, j, b)$ \Comment{top-$b$ among all $\Qsf{S}{q_i}{q_j}$}

    \EndFor    \State \Return $\Qsf{S}{q_0}{q_n}$ 
\EndFunction
\end{algorithmic}
\end{minipage} 
\begin{minipage}{.54\textwidth}
\begin{algorithmic}
\Function{LD-ExPartition}{{\color{red}$\distri = \langle D, \tau \rangle$}, $b$} \Comment{$b$ is beam size}
    \State $n \gets$ length of $D$
    \State $\expQ^{S} \gets$ hash() \Comment{hash table: from state $[q_i, q_j]$ to $\expQsf{S}{q_i}{q_j}$}
    \State $\expQ^{P} \gets$ hash() \Comment{hash table: from state $[q_i, q_j]$ to $\expQsf{P}{q_i}{q_j}$}
    \For{$j=4 \ldots n$}
      \For {$q_{j-1} \in \nodes(D, j-1)$ and $q_i$ s.t. $(q_i, q_{j-1}) \in \expQ^S$}
         	\For {$\longlaedgew{q_{j-1}}{b:\textcolor{red}{\tau(q_{j-1}, b)}}{q_j} \in \outedges(D, q_{j-1})$} 
            \State $\expQsf{S}{q_i}{q_j} \pluseq \textcolor{red}{\tau(q_{j-1}, b) \,\cdot\,}  \expQsf{S}{q_i}{q_{j-1}} \cdot e^{-\frac{\delta(b)}{RT}} $
		     \For {$\longlaedgew{q_{i-1}}{a:\textcolor{red}{\tau(q_{i-1}, a)}}{q_i} \in \inedges(D, q_i)$} 
                       \If {$a, b$ in \allpairtypes}
                       \State $\expQsf{P}{q_{i-1}}{q_j} \pluseq \textcolor{red}{\tau(q_{i-1}, a)\cdot \tau(q_{j-1}, b) \,\cdot\,} \expQsf{S}{q_i}{q_{j-1}} \cdot e^{-\frac{\xi(a, b)}{RT}}$ 
                     \EndIf

		   \EndFor
	        \EndFor
	     \EndFor
      \State \textproc{BeamPrune}$(\expQ^S, \expQ^P, j, b)$ \Comment{top-$b$ among all $\expQsf{P}{q_i}{q_j}$}
      \For {$q_j \in \nodes(D, j)$}
         \For {$q_i$ such that $(q_i, q_j) \in \expQ^P$}
            \For {$q_k$ such that $(q_k, q_i) \in \expQ^S$}
                \State $\expQsf{S}{k}{j} \pluseq \expQsf{S}{k}{i} \cdot \expQsf{P}{i}{j}$ 

           \EndFor          
         \EndFor
      \EndFor
      \State \textproc{BeamPrune}$(\expQ^S, \expQ^S, j, b)$ \Comment{top-$b$ among all $\expQsf{S}{q_i}{q_j}$}

    \EndFor    \State \Return $\expQsf{S}{q_0}{q_n}$  
\EndFunction
\end{algorithmic}
\end{minipage}
\caption{
The pseudocode of (simplified) partition function and expected partition function calculations based on LinearDesign algorithm. 
\label{fig:linear}}
\end{figure*}

%% file: fig_beam_pruning.tex

\begin{figure}[h]
\hspace{-10pt}
\small
\begin{minipage}{1.05\linewidth}
\begin{algorithmic}
\Function{BeamPrune}{$Q^S, Q^X, j, b$}
    \State $\cands \gets$ hash() \Comment{hash table: from $q_i$ to score $\Qsf{S}{q_0}{q_i} + \Qsf{X}{q_i}{q_j}$}
    \For{$q_j \in \nodes(j)$}s
	\For{$(q_i, q_j) \in Q^X$}
        	      \State $\cands[q_i] \gets \Qsf{S}{q_0}{q_i} + \Qsf{X}{q_i}{q_j}$ \Comment{$\Qsf{S}{q_0}{q_i}$ as prefix score}
        \EndFor
    \EndFor
    \State $\cands \gets \textproc{SelectTopB}(\cands, b)$ \Comment{select top-$b$ by score}
    \For{$(q_i, q_j)  \in Q^X$}
        \If{$q_i \not\in \cands$}
            \State delete $(q_i, q_j)$ in $Q^X$ \Comment{prune out low-scoring states}
        \EndIf
    \EndFor
\EndFunction 
\end{algorithmic}
\end{minipage}
\caption{
The pseudocode of beam pruning. 
\label{fig:beam_pruning}}
\end{figure}

%% file: DFA_formulation.tex

In the context of mRNA design challenges, our objective is to derive an mRNA sequence, $\vec{x} = (x_1, x_2, \ldots, x_{3n})$, where $x_i$ represents the $i$-th nucleotide, corresponding to a given target protein sequence, $\vec{p} = (p_1, p_2, \ldots, p_n)$ with $p_i$ indicating the $i$-th amino acid residue. This mRNA sequence must be capable of being translated into the target protein. Following the framework proposed in LinearDesign~\cite{zhang+:2023}, we can model all feasible candidate mRNA sequences through a representation known as an mRNA Deterministic Finite Automaton (DFA).

A DFA is formally defined as a 5-tuple $D = \langle Q, \Sigma, \delta, q_0, F \rangle$, where:
\begin{itemize}
\item $\; Q$ denotes the set of states,
\item $\; \Sigma$ is the alphabet, for our purposes $\Sigma=\{\text{A}, \text{C}, \text{G}, \text{U}\}$,
\item $\; q_0$ is the initial state, denoted as $\text{state}(0,0)$ in our analysis,
\item $\; F$ represents the set of final states, which is unique in this context,
\item $\; \delta$ is the transition function that maps a state $q$ and a symbol $a \in \Sigma$ to a subsequent state $q'$, symbolically, $\delta(q, a) = q'$, which is represented as a labeled edge from $q$ to $q'$ with label $a$.
\end{itemize}

To construct the mRNA DFA, we begin by creating individual DFAs for each amino acid. These are then concatenated to form a single comprehensive DFA, $D(\vec{p})$, for the entire protein sequence $\vec{p}$. This collective DFA represents all possible mRNA sequences that could translate into the protein:
\[D(\vec{p}) = D(p_0) \circ D(p_1) \circ \cdots \circ D(p_{|\vec{p}|-1}) \circ D(\text{stop})\]
where the end state of the complete mRNA DFA is $\text{state}(3|\vec{p}|+3, 0)$, and the length of the DFA is defined as $3|\vec{p}|+3$.

Extending this framework, we introduce a probabilistic DFA (pDFA) defined by a 6-tuple $\distri = \langle D, \tau \rangle = \langle Q, \Sigma, \delta, q_0, F, \tau \rangle$, where the initial five components are identical to those in $D$, and $\tau$ represents the probability function. This function assigns to each transition from state $q \in Q$ using symbol $a \in \Sigma$ a probability within the range $[0, 1]$. Specifically, for any state $q$, the probability distribution is given by:
\[
\tau(q, \cdot) : \nucset(q) \mapsto [0,1], \text{ s.t. } \sum_{x \in \nucset(q)} \tau(q, x) = 1.
\]
where $\nucset(q)$ denotes the set of nucleotides allowed from state $q$, as determined by the transition function $\delta$, defining the valid transitions from $q$ with nucleotide $a$. Thus, each path through $\distri$ is associated with a probability defined by $\tau$, making $\distri$ a distribution over possible mRNA sequences. 

In the pDFA \distri, the probability of an mRNA sequence $\vecx$ within the distribution defined by it is conceptualized as the product of probabilities across each node along the sequence's path:
\[
\distri(\vecx) = \prod_{i} \tau(q_i, x_{i}), \; \text{given} \; q_0 = (0,0) \; \text{and} \; q_{i+1} = \delta(q_i, x_i),
\]
where $q_0 = \text{state}(0,0)$ is the starting node, and $q_{i+1} = \delta(q_i, x_i)$  is the next node reached by following the edge corresponding to nucleotide $x_i$.

For the sake of clarity in pseudocode representation, we define $\text{nodes}(D, j) = \{(j,0), (j, 1) \}$ as the set of nodes at position $j$, $\text{in\_edges}(q) = \{\text{edge}(q',a,q) \mid \delta(q', a) = q \}$ to denote the set of incoming edges to state $q$, and $\text{out\_edges}(q) = \{\text{edge}(q,a,q') \mid \delta(q, a) = q' \}$ to denote the set of outgoing edges from state $q$.

%% file: main.bbl
\begin{thebibliography}{}

\bibitem[Baden {\em et~al.}(2021)Baden, El~Sahly, Essink, Kotloff, Frey, Novak,
  Diemert, Spector, Rouphael, Creech, {\em et~al.}]{baden+:2021}
Baden, L.~R.  {\em et~al.} (2021).
\newblock Efficacy and safety of the m{RNA}-1273 {SARS-CoV-2} vaccine.
\newblock {\em New England Journal of Medicine\/}, {\bf 384}(5), 403--416.

\bibitem[Bonnet {\em et~al.}(2020)Bonnet, Rzazewski, and
  Sikora]{bonnet2020designing}
Bonnet, {\'{E}}.  {\em et~al.} (2020).
\newblock {Designing RNA secondary structures is hard}.
\newblock {\em Journal of Computational Biology\/}, {\bf 27}(3), 302--316.

\bibitem[Cohen and Skiena(2003)Cohen and Skiena]{cohen+skiena:2003}
Cohen, B. and Skiena, S. (2003).
\newblock Natural selection and algorithmic design of m{RNA}.
\newblock {\em Journal of Computational Biology\/}, {\bf 10}(3-4), 419--432.

\bibitem[Consortium(2022)Consortium]{UniProt}
Consortium, T.~U. (2022).
\newblock {UniProt: the Universal Protein Knowledgebase in 2023}.
\newblock {\em Nucleic Acids Research\/}, {\bf 51}(D1), D523--D531.

\bibitem[Do {\em et~al.}(2006)Do, Woods, and Batzoglou]{do+:2006}
Do, C.~B.  {\em et~al.} (2006).
\newblock Contrafold: {RNA} secondary structure prediction without
  physics-based models.
\newblock In {\em Proceedings 14th International Conference on Intelligent
  Systems for Molecular Biology 2006, Fortaleza, Brazil, August 6-10, 2006\/},
  pages 90--98.

\bibitem[Eisner(2016)Eisner]{eisner:2016}
Eisner, J. (2016).
\newblock Inside-outside and forward-backward algorithms are just backprop
  (tutorial paper).
\newblock In K.~Chang, M.~Chang, A.~M. Rush, and V.~Srikumar, editors, {\em
  Proceedings of the Workshop on Structured Prediction for NLP@EMNLP 2016,
  Austin, TX, USA, November 5, 2016\/}, pages 1--17. Association for
  Computational Linguistics.

\bibitem[Huang {\em et~al.}(2019)Huang, Zhang, Deng, Zhao, Liu, Hendrix, and
  Mathews]{huang+:2019}
Huang, L.  {\em et~al.} (2019).
\newblock {LinearFold: linear-time approximate {RNA} folding by 5'-to-3'
  dynamic programming and beam search}.
\newblock {\em Bioinformatics\/}, {\bf 35}(14), i295--i304.

\bibitem[Lorenz {\em et~al.}(2011)Lorenz, Bernhart, zu~Siederdissen, Tafer,
  Flamm, Stadler, and Hofacker]{lorenz+:2011}
Lorenz, R.  {\em et~al.} (2011).
\newblock Viennarna package 2.0.
\newblock {\em Algorithms Mol. Biol.}, {\bf 6}, 26.

\bibitem[Matthies {\em et~al.}(2023)Matthies, Krueger, Torda, and
  Ward]{matthies+:2023}
Matthies, M.  {\em et~al.} (2023).
\newblock {Differentiable Partition Function Calculation for RNA}.
\newblock {\em Nucleic Acids Research\/}.

\bibitem[Mauger {\em et~al.}(2019)Mauger, Cabral, Presnyak, Su, Reid, Goodman,
  Link, Khatwani, Reynders, Moore, {\em et~al.}]{mauger+:2019}
Mauger, D.~M.  {\em et~al.} (2019).
\newblock m{RNA} structure regulates protein expression through changes in
  functional half-life.
\newblock {\em Proceedings of the National Academy of Sciences {U.S.A.}}, {\bf
  116}(48), 24075--24083.

\bibitem[McCaskill(1990)McCaskill]{mccaskill:1990}
McCaskill, J.~S. (1990).
\newblock The equilibrium partition function and base pair probabilities for
  {RNA} secondary structure.
\newblock {\em Biopolymers\/}, {\bf 29}, 1105--19.

\bibitem[Polack {\em et~al.}(2020)Polack, Thomas, Kitchin, Absalon, Gurtman,
  Lockhart, Perez, P{\'e}rez~Marc, Moreira, Zerbini, {\em
  et~al.}]{polack+:2020}
Polack, F.~P.  {\em et~al.} (2020).
\newblock {Safety and efficacy of the BNT162b2 mRNA Covid-19 vaccine}.
\newblock {\em New England journal of medicine\/}, {\bf 383}(27), 2603--2615.

\bibitem[Terai {\em et~al.}(2016)Terai, Kamegai, and Asai]{terai+:2016}
Terai, G.  {\em et~al.} (2016).
\newblock {CDSfold}: an algorithm for designing a protein-coding sequence with
  the most stable secondary structure.
\newblock {\em Bioinformatics\/}, {\bf 32}(6), 828--834.

\bibitem[Turner and Mathews(2009)Turner and Mathews]{turner+:2009}
Turner, D.~H. and Mathews, D.~H. (2009).
\newblock {NNDB: the nearest neighbor parameter database for predicting
  stability of nucleic acid secondary structure}.
\newblock {\em Nucleic Acids Research\/}, {\bf 38}, D280--D282.

\bibitem[Wikipedia(2023)Wikipedia]{wikipedia2023codonwh}
Wikipedia (2023).
\newblock {DNA and RNA codon tables}.
\newblock Wikipedia, The Free Encyclopedia.

\bibitem[Wu {\em et~al.}(2020)Wu, Zhao, Yu, Chen, Wang, Song, Hu, Tao, Tian,
  Pei, {\em et~al.}]{wu+:2020:sarscov2}
Wu, F.  {\em et~al.} (2020).
\newblock A new coronavirus associated with human respiratory disease in china.
\newblock {\em Nature\/}, {\bf 579}(7798), 265--269.

\bibitem[Zhang {\em et~al.}(2020)Zhang, Zhang, Mathews, and
  Huang]{zhang+:2020linearpartition}
Zhang, H.  {\em et~al.} (2020).
\newblock {LinearPartition: linear-time approximation of RNA folding partition
  function and base-pairing probabilities}.
\newblock {\em Bioinformatics\/}, {\bf 36}(Supplement\_1), i258--i267.

\bibitem[Zhang {\em et~al.}(2023)Zhang, Zhang, Lin, Xu, Li, Liu, Liu, Ma, Zhao,
  Jiang, Chen, Shen, Li, Mathews, Zhang, and Huang]{zhang+:2023}
Zhang, H.  {\em et~al.} (2023).
\newblock {Algorithm for optimized mRNA design improves stability and
  immunogenicity}.
\newblock {\em Nature\/}, {\bf 621}(7978), 396--403.

\bibitem[Zhou {\em et~al.}(2023)Zhou, Dai, Li, Ward, Mathews, and
  Huang]{zhou+:2023samfeo}
Zhou, T.  {\em et~al.} (2023).
\newblock {RNA design via structure-aware multifrontier ensemble optimization}.
\newblock {\em Bioinformatics\/}, {\bf 39}(Supplement\_1), i563--i571.

\end{thebibliography}
